\documentclass[prl, pdflatex, twocolumn, superscriptaddress, notitlepage, floatfix]{revtex4-2}
\usepackage{xcolor}
\usepackage{graphicx}
\usepackage{wrapfig}
\usepackage{amsmath}
\usepackage{amssymb}
\usepackage{multirow}
\usepackage{slashed}
\usepackage{physics}
\usepackage{soul}
\setstcolor{red}
\usepackage[colorlinks=true, linkcolor=blue, citecolor=blue, urlcolor=blue]{hyperref}

\newcommand{\m}{\mu_B}
\newcommand{\Dn}[1]{\bar{D}_{#1}}

\newcommand{\Gn}[1]{\bar{\cal G}_{#1}}
\newcommand{\cb}{\chi^B}
\newcommand{\av}[1]{\left\langle{#1}\right\rangle}
\newcommand{\pt}{\Delta P^E}
\newcommand{\pr}{\Delta P^R}
\newcommand{\ndt}{\mathcal{N}^E}
\newcommand{\ndr}{\mathcal{N}^R}
\newcommand{\zr}{Z^R}
\newcommand{\thr}{\Theta^R}
\newcommand{\tht}{\Theta^E}
\newcommand{\nth}[1]{{#1}^\text{th}}

\newcommand{\dbeta}{\Delta\beta}
\newcommand{\dT}{\Delta T}
\newcommand{\dm}{\Delta m}
\newcommand{\hm}{\hat{\mu}_B}
\newcommand{\hms}{\hat{m}}
\newcommand{\hdm}{\Delta\hat{m}}

\begin{document}
\title{Lattice QCD Equation of State for Nonvanishing Chemical Potential by Resumming
Taylor Expansion}
\author{Sourav Mondal}
\affiliation{Centre for High Energy Physics, Indian Institute of Science, Bangalore 560012, India.}
\author{Swagato Mukherjee}
\affiliation{Physics Department, Brookhaven National Laboratory, Upton, New York 11973, USA.}
\author{Prasad Hegde}
 \email{prasadhegde@iisc.ac.in}
 \affiliation{Centre for High Energy Physics, Indian Institute of Science, Bangalore 560012, India.}
\begin{abstract}
Taylor expansion in powers of baryon chemical potential ($\m$) is an oft-used
method in lattice QCD to compute QCD thermodynamics for $\m>0$. Based only upon
the few known lowest order Taylor coefficients, it is difficult to discern the
range of $\m$ where such an expansion around $\m=0$ can be trusted. We introduce
a resummation scheme for the Taylor expansion of the QCD equation of state in
$\m$  that is based on the $n$-point correlation functions of the conserved
current ($D_n$).  The method resums the contributions of the first $N$
correlation function $D_1$, $\dots$, $D_N$ to the Taylor expansion of the QCD
partition function to all orders in $\m$. We show that the resummed partition
function is an approximation to the reweighted partition function at $\m\ne0$.
We apply the proposed approach to high-statistics lattice QCD calculations using
2+1 flavors of Highly Improved Staggered Quarks with physical quark masses on
$32^3\times8$ lattices and for temperatures $T\approx145-176$~MeV. We
demonstrate that, as opposed to the Taylor expansion, the resummed version not
only leads to improved convergence but also reflects the zeros of the resummed
partition function and severity of the sign problem, leading to its eventual
breakdown. We also provide a generalization of our scheme to include resummation of
powers of temperature and quark masses in addition to $\m$, 
and show that the alternative expansion scheme of
[S. Borsányi et al., Phys. Rev. Lett. {\bf 126}, 232001 (2021).] 
is a special case of this generalized resummation.
\end{abstract}
\date{\today}
\maketitle
%
\emph{Introduction.--} Lattice Quantum Chromodynamics (QCD) results for the QCD
equation of state (EoS) plays a critical role in the dynamical modeling of heavy-ion
collisions~\cite{Bernhard:2016tnd, Parotto:2018pwx, Monnai:2019hkn, Everett:2020xug}
and, thereby, in the experimental explorations of the QCD phase diagram in the
$T$-$\m$ plane. Due to the fermion sign problem it is difficult to carry out lattice
QCD computations directly at $\m\ne0$. Despite some recent
progress~\cite{Cristoforetti:2012su, Sexty:2013ica, Fukuma:2019uot, Aarts:2009yj,
Aarts:2013uxa, Fodor:2015doa}, direct lattice computations of the QCD EoS $\m\ne0$
with physical quark masses, fine lattice spacings and large lattice volumes have remained
elusive. Instead, the present state-of-the-art lattice QCD EoS at $\m>0$ has been
obtained using the Taylor expansion~\cite{Bazavov:2017dus, Datta:2016ukp} and the
analytic continuation~\cite{Borsanyi:2018grb, Gunther:2016vcp} methods. In the Taylor
expansion method one expands the pressure in powers of $\m$ around $\m=0$ and
directly computes the Taylor coefficients at $\m=0$. For the analytic continuation,
one avoids the fermion sign problem using simulations at purely imaginary values
$\m$, fits these results with a power series in $\m$ to determine the Taylor
coefficients at $\m=0$ and then provides the EoS at real $\m>0$ based on these Taylor
coefficients. On the other hand, it is well-known that the applicability of the Taylor
expansion as well as the analytic continuation should be limited by the zeros,
nearest to $\m=0$, of the partition function in the entire complex-$\m$
plane~\cite{Stephanov:2006dn, Almasi:2019bvl, Mukherjee:2019eou}. In principle, it is
possible to  gain some knowledge about the locations of the zeros of the partition
function by re-expressing the power series in real or imaginary $\m$ in terms of
Pad\'e approximants~\cite{Datta:2016ukp} or in a power series of the
fugacity~\cite{Barbour:1991vs, Giordano:2019gev, Giordano:2019slo}. Armed, in reality, with only the few
lowest order Taylor coefficients, this becomes a very difficult task and, in practice,
one just restricts the EoS to $\{T,\m\}$ that avoids any pathological nonmonotonicity
in the truncated Taylor series~\cite{Bazavov:2017dus, Gunther:2016vcp}. Furthermore,
these methods provide very little guidance on the severity of the fermion problem,
\textit{i.e.} how rapidly the phase of the partition function fluctuates as $\m$ is
increased.  It is possible to determine the zeros of the partition function as well
as its average phase by reweighting the fermion determinant to
$\m\ne0$~\cite{Fodor:2001au, Fodor:2004nz, Ejiri:2004yw, Saito:2013vja,
Giordano:2020roi}. However, due to the computational cost associated with exact
evaluation of the fermion determinant, at present this method is restrained within
coarse lattice spacings and small lattice volumes.

In this work, we introduce a method for the calculation of the lattice QCD EoS that
genuinely resums the truncated Taylor series to all orders in $\m$ and whose
breakdown encodes the severity of the sign problem and zeros of the resummed
partition function.

%
%
\emph{The resummation method.--} The Taylor expansion to $\order{\m^N}$ of the excess
pressure, $\Delta P(T,\m) \equiv P(T,\m)-P(T,0)$, is given by
\begin{align}
  \frac{\pt_N}{T^4} = \sum_{n=1}^N \frac{\cb_n}{n!} \qty(\frac{\m}{T})^n ,
  \label{eq:pt}
\end{align}
where the Taylor coefficients are defined as
\begin{align}
  \cb_n(T) = \frac{1}{VT^3} \left. \pdv[n]{\ln Z(T,\m)}{(\m/T)} \right\vert_{\m=0} .
  \label{eq:chi}
\end{align}
Here, the QCD partition function is denoted as $Z = \int e^{-S} \det[M]
\mathcal{D}U$, $V$ is the spatial volume, $U$ is the $SU(3)$ gauge fields,  $S$ is
the pure gauge action and $M$ is the fermion matrix. Each $\cb_n$ consists of sum
of terms like $\av{D_i^a D_j^b \cdots D_k^c}$ with $i \cdot a + j \cdot b + \cdots +
k \cdot c=n$~\cite{Allton:2005gk,Gavai:2003mf}, where
\begin{align}
  D_n(T) = \bar{D}_n\cdot n! = \left. \pdv[n]{\ln\det[M(T,\m)]}{(\m/T)} \right\vert_{\m=0} ,
  \label{eq:D}
\end{align}
and the $\av{\cdot}$ denotes average over gauge field ensembles at $\m=0$,
\textit{i.e.}  $\av{O}=\int O e^{-S} \det[M(T,0)] \mathcal{D}U/Z$. The physical
interpretation of $D_n$ is simple for the continuum theory:  $D_n=\int \dd\vb{x_1}
\cdots \dd\vb{x_n} J_0(\vb{x_1}) \cdots J_0(\vb{x_n})$  is the integrated $n$-point
correlation function of the $\nth{0}$ component of the conserved (baryon)
current $J_0(\vb{x})$ at a space-time point $\vb{x}$. Note that, due to $CP$ symmetry
of QCD all $D_n$ for $n=odd(even)$ are purely imaginary(real) and only the $n=even$
terms contribute to \autoref{eq:pt}. In practice, lattice QCD computations of the
$\cb_N$ involve computations of all $D_n$ for $n\le N$ as intermediate steps, and
$\cb_N$ are obtained from combinations of $D_n$ and their powers.

\begin{figure}[!t]
  \centering
  \includegraphics[width=0.45\textwidth, height=0.22\textheight]{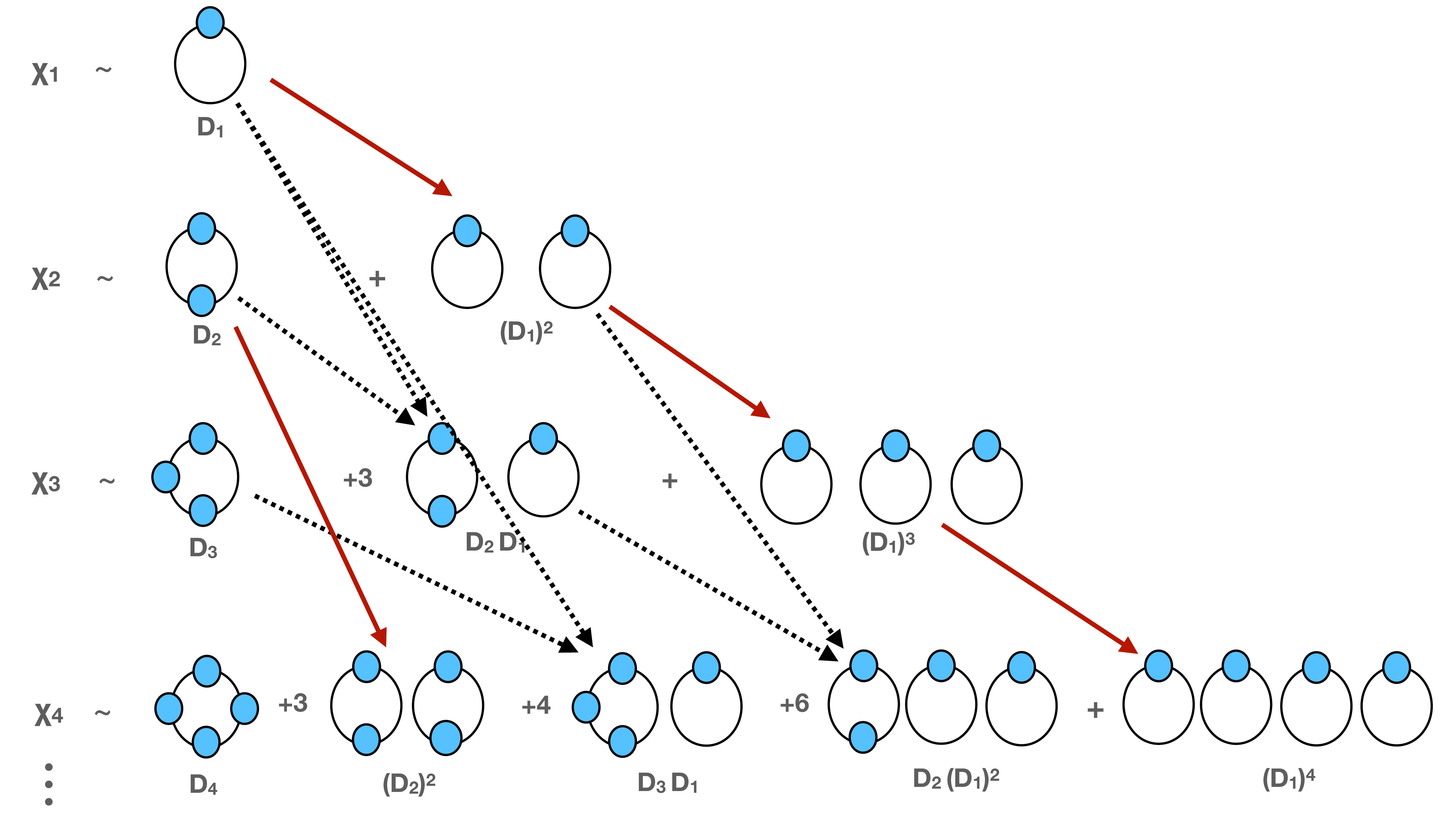}
  \caption{Contributions of different $D_n$ to the $\cb_n$. Each blob represents
  insertion of the $\nth{0}$ component of the conserved current. Solid red and
  dotted black lines represent directly exponentiated and cross terms respectively.}
  \label{fig:D}
\end{figure}

Contributions of various combinations of $D_n$ to the few lowest order Taylor
coefficients are sketched in \autoref{fig:D}. If one considers the factorials and the
powers of $\m/T$ associated with each $D_n$ in the sum of \autoref{eq:pt}, it is not
difficult to realize that all contributions of each $D_n$ to $\pt$ can be resummed
into exponential forms. For example, contributions of $D_1^n$ from all $\cb_n$ in
\autoref{eq:pt} can be resummed as $\exp[\bar{D}_1(\m/T)]$. Similarly, contributions of all
$D_2^n$ can be resummed as $\exp[\bar{D}_2(\m/T)^2]$, and so on. Also it is  easy to see
that the contributions of the mixed terms like $D_1^nD_2^m$ arise from
$\exp[\bar{D}_1(\m/T)] \times \exp[\bar{D}_2(\m/T)^2]$. Thus, it is possible to write down a
resummed version of \autoref{eq:pt}, \textit{viz.}
\begin{align}
  \frac{\pr_N}{T^4}  = \frac{1}{T^3V} \ln \av{ \exp[ \sum_{n=1}^N \bar{D}_n \qty(\frac{\m}{T})^n ] } ,
  \label{eq:pr}
\end{align}
providing the EoS up to infinite orders in $\m$.  The $\pr_N$ can be considered as a
$\m$-dependent effective action obtained by resumming up to $N$-point correlation
functions of the conserved current. Expansion of $\pr_N$ in powers of $\m/T$ yields
an infinite series in $\m/T$, in addition to the truncated Taylor series: $\pt_N +
\sum_{n>N}^\infty \langle \bar{D}_1^i \cdots \bar{D}_N^j \rangle (\m/T)^n$, where $i, j = 0,
\dots, N$ satisfying $1 \cdot i + \cdots + N \cdot j = n$.
The Taylor expanded ($\ndt_N$) and the resummed ($\ndr_N$) net baryon-number
densities can be straightforwardly obtained as a single $\m$-derivative of
$\pt$ and $\pr$ in \autoref{eq:pt} and \autoref{eq:pr}, respectively.

The resummed version in \autoref{eq:pr} also highlights the connection between  the
Taylor expansion and the reweighting method. In the reweighting method
$Z(T,\m)/Z(T,0) = \langle \det[M(T,\m)]/\det[M(T,0)] \rangle$ can be calculated, if
computationally feasible, by exactly evaluating the ratio of the fermion matrix
determinants on the gauge fields generated at $\m=0$. In more realistic lattice
calculations with large volumes, exact evaluations of the determinant ratios might not
be computationally feasible and one may consider evaluating $\det[M(T,\m)]$ within
some approximation scheme to obtain approximate partition function
$\zr_N(T,\m)\approx Z(T,\m)$. Following the spirit of the Taylor expansion, one such
approximation scheme can be expansion of $\det[M(T,\m)]$ in powers of $\m/T$. Keeping
in mind $\det[M] = \exp[\Tr\ln M]$ and \autoref{eq:D}, one can immediately recognize
\begin{align}
   \frac{\zr_N(T,\m)}{Z(T,0)} = \av{  \exp[ \sum_{n=1}^N \bar{D}_n \qty(\frac{\m}{T})^n ] } .
  \label{eq:Z}
\end{align}
Since $CP$ symmetry dictates that the even(odd) $D_n$ are purely real(imaginary) and
the partition function must be real, a measure of the severity of the sign problem is
given by the average phase factor for $\zr_N$ (with $\m$ real),
\begin{align}
  \av{\cos\thr_N} = \av{ \cos(\sum_{n=1}^{N/2} \Im[\bar{D}_{2n-1}] \qty(\frac{\m}{T})^{2n-1} ) }.
  \label{eq:theta}
\end{align}
An expansion of $\av{\cos\thr_N}$ in $\m/T$ leads to the Taylor expanded measure of the
average phase of the partition function~\cite{Ejiri:2004yw, Allton:2005gk}, which we
will denote by $\tht_N$. As the sign problem becomes more severe the average phase
$\av{\cos\thr_N}\approx0$ and resummed results will also show signs of breakdown.
Furthermore, although $\pt_N$ 
can be evaluated for any complex value of $\m$, $\pr_N$ becomes undefined when $\Re[\zr_N]\le0$ for a given $N$ and
statistics, leading to a natural breakdown of the resummed results.
The location of
the zeros of $\zr_N$ in the complex-$\m$ plane will indicate the $\m$ region
where such resummation can be applicable. Obviously, for any given $N$ the region of
applicability of $\pt_N$ cannot exceed the same for $\pr_N$.

%
%

\emph{Lattice QCD computations.--} For this work, we used the data for $\cb_n$ and
$D_n$ generated by the HotQCD collaboration for calculations of the QCD
EoS~\cite{Bazavov:2017dus} and the chiral crossover
temperature~\cite{Bazavov:2018mes} at $\m>0$ using the Taylor expansion method.  The
HotQCD ensembles were generated with 2+1-flavors of Highly Improved Staggered
Quarks and the tree-level improved Symanzik gauge action~\cite{Bazavov:2011nk,
Bazavov:2012jq, Bazavov:2014pvz}. Bare quark masses were chosen to reproduce, within
a few percent, the physical value of the kaon mass and a pseudo-Goldstone pion mass
of $138$~MeV in the continuum limit at $T=\m=0$ and the lattice spacing were
calibrated against the physical value of the kaon decay constant
~\cite{Bazavov:2019www}. We present lattice QCD results from a single lattice size
$32^3\times8$ and for 6 temperatures  $T= 145, 151, 157, 166, 171, 176$~MeV.
About 475K, 520K, 716K, 522K, 232K and 152K gauge field configurations were
used to measure $D_n$ at these temperatures respectively.  The gauge field
configurations were separated by 10 Rational Hybrid Monte Carlo trajectories of unit
length. The $D_n$ were calculated within the formalism adopted in
Refs.~\cite{Bazavov:2017dus, Bazavov:2018mes}, \textit{i.e.} using the exponential-$\mu$
formalism~\cite{Hasenfratz:1983ba} for $n\le4$ and the linear-$\mu$
formalism~\cite{Gavai:2011uk, Gavai:2014lia} for $n>4$. The expressions for $D_n$ in terms
of the traces involving the inverse of the staggered fermion matrix and its $\m$-derivatives
are well-known~\cite{Allton:2005gk,Steinbrecher:2018jbv}. Each trace was calculated stochastically
for each configuration by employing 2000 random Gaussian volume sources for the trace $D_1$
and 500 random sources for the rest~\cite{Steinbrecher:2018jbv}.

%
%

%
\begin{figure}[!t]
  \centering
  \includegraphics[width=0.4\textwidth]{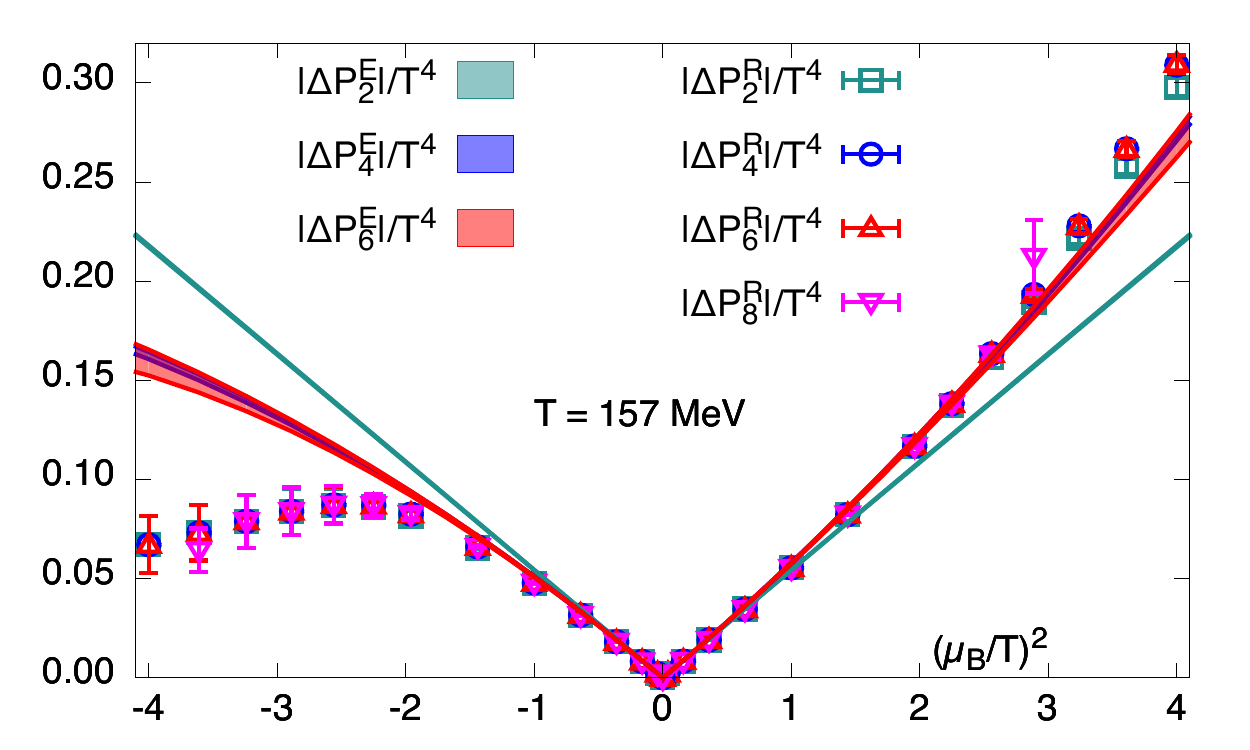}
  \includegraphics[width=0.4\textwidth]{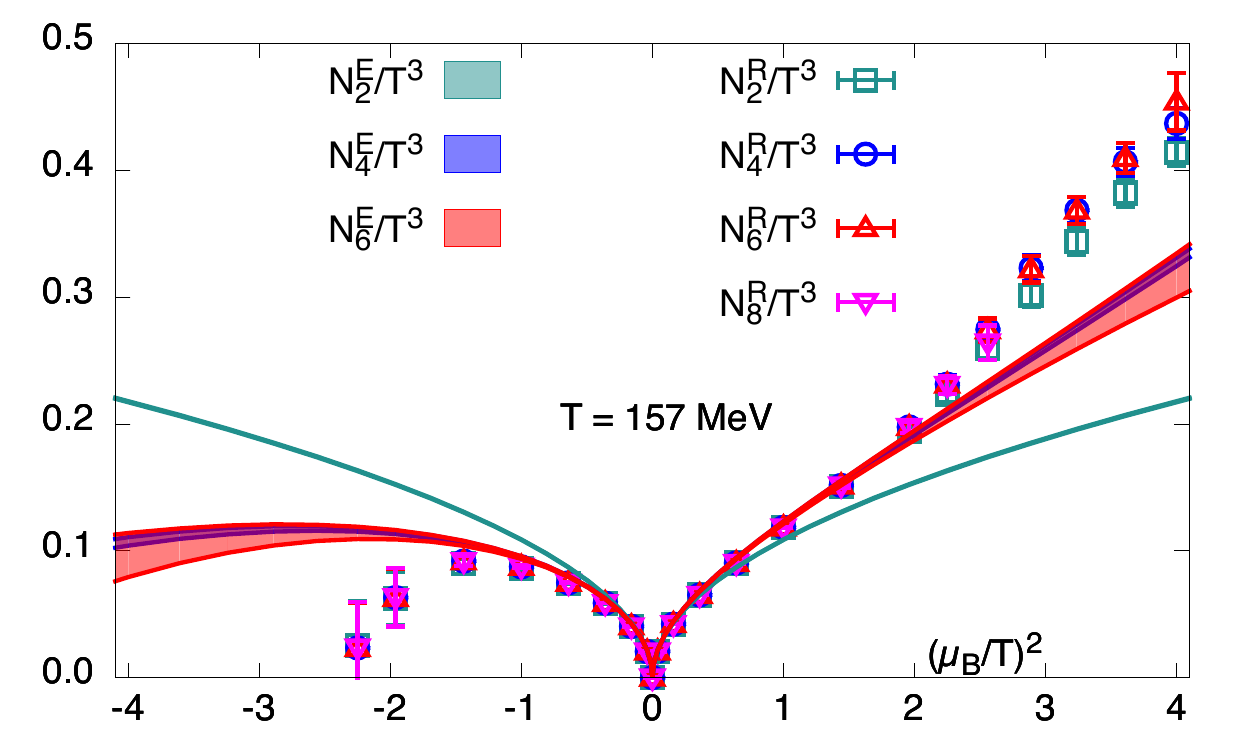}
  \caption{Comparisons between the Taylor expanded and resummed results for different orders
  for  the excess pressure
  (top) and net baryon-number density (bottom) at $T=157$~MeV. Results for real and
  imaginary $\m/T$ are plotted on the positive and negative $x$-axis respectively.}
  \label{fig:results_b6390}
\end{figure}

\emph{Results.--} To demonstrate the superiority of the resummation method over the
Taylor expansion, we chose the temperature where we had the largest statistics,
\textit{i.e.} $T=157$~MeV, which is also closest to the QCD crossover
temperature~\cite{Bazavov:2018mes}. In Fig.~\ref{fig:results_b6390}, we compare
$\pt_N$ with $\pr_N$ (top) and $\ndt_N$ with $\ndr_N$ for different orders $N$.
Comparisons are shown both for real as well as imaginary values of $\m$, corresponding
to positive and negative values of $(\mu_B/T)^2$, respectively. The $\pr_N$ and
$\ndr_N$ show very good convergence between different orders $N=2, 4, 6, 8$. The
Taylor-expanded results seem to approach their respective resummed results as
contributions from higher orders in $\m$ are included; however the convergence of the
Taylor-expanded results is slow due the alternating signs of the higher order
$\cb_n$ near the QCD crossover~\cite{Bazavov:2017dus, Datta:2016ukp,
Borsanyi:2018grb}. The resummation method overcomes this problem by including
contributions from all orders in $\m$ and shows markedly improved convergence. In
contrast to the Taylor expansion, the resummed results break down for
 $\lvert\m/T\gtrsim1.5\rvert$.
For $\lvert\m/T\gtrsim1.5\rvert$, $\Re[Z_8]\le0$ and $\ndr_8$ becomes
divergingly large. We checked that such a breakdown is not a mere statistical issue by
repeating the calculations using only parts of the gauge configurations available at
this temperature. Similar breakdown for $\m/T\gtrsim1.5$ was also observed in
Refs.~\cite{Datta:2016ukp, Gupta:2014qka, Gavai:2008zr} when the EoS was
reconstructed from the Pad\'e approximants of the Taylor series in $\m$. While
Pad\'e-based continuations of the QCD crossover temperature from imaginary values
$\m$ did not encounter such breakdowns~\cite{Cea:2012ev, Pasztor:2020dur}, the same in the case of
the EoS seemed to break down due to singularities in the complex-$\m$ plane~\cite{Schmidt:2021pey}.

\begin{figure}[!t]
  \centering
  \includegraphics[width=0.4\textwidth]{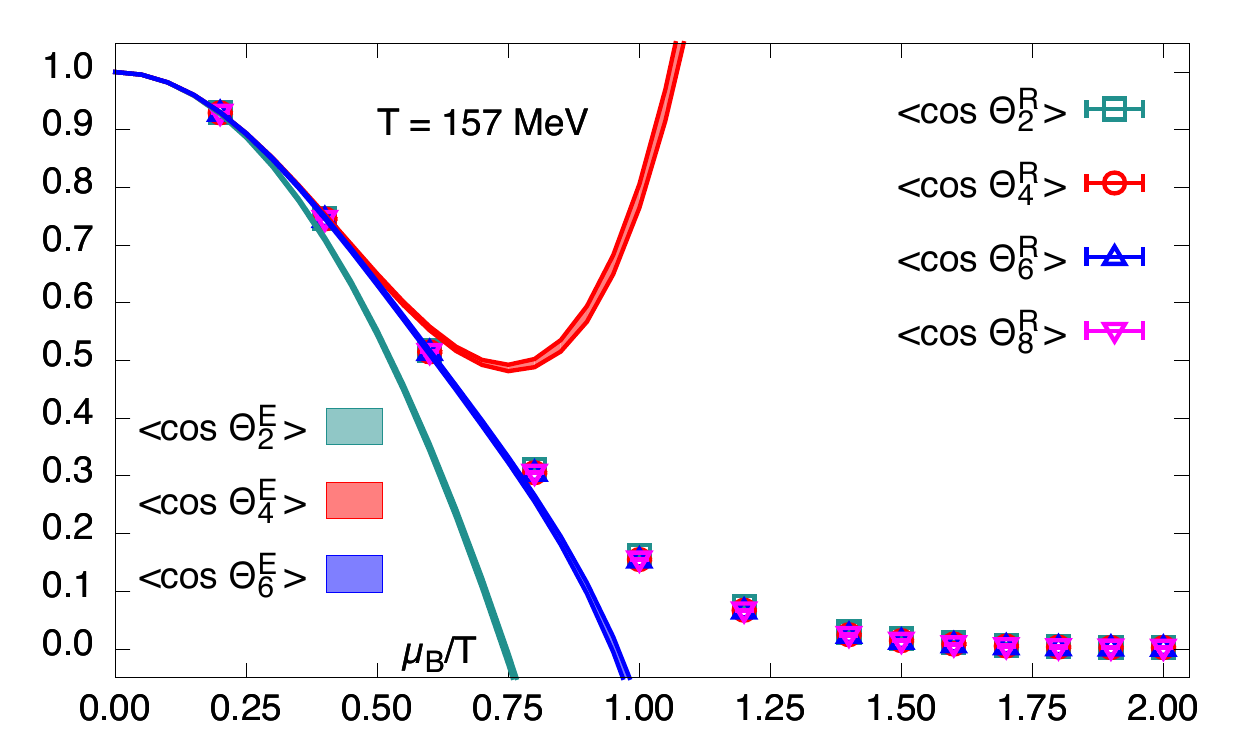}
  \includegraphics[width=0.4\textwidth]{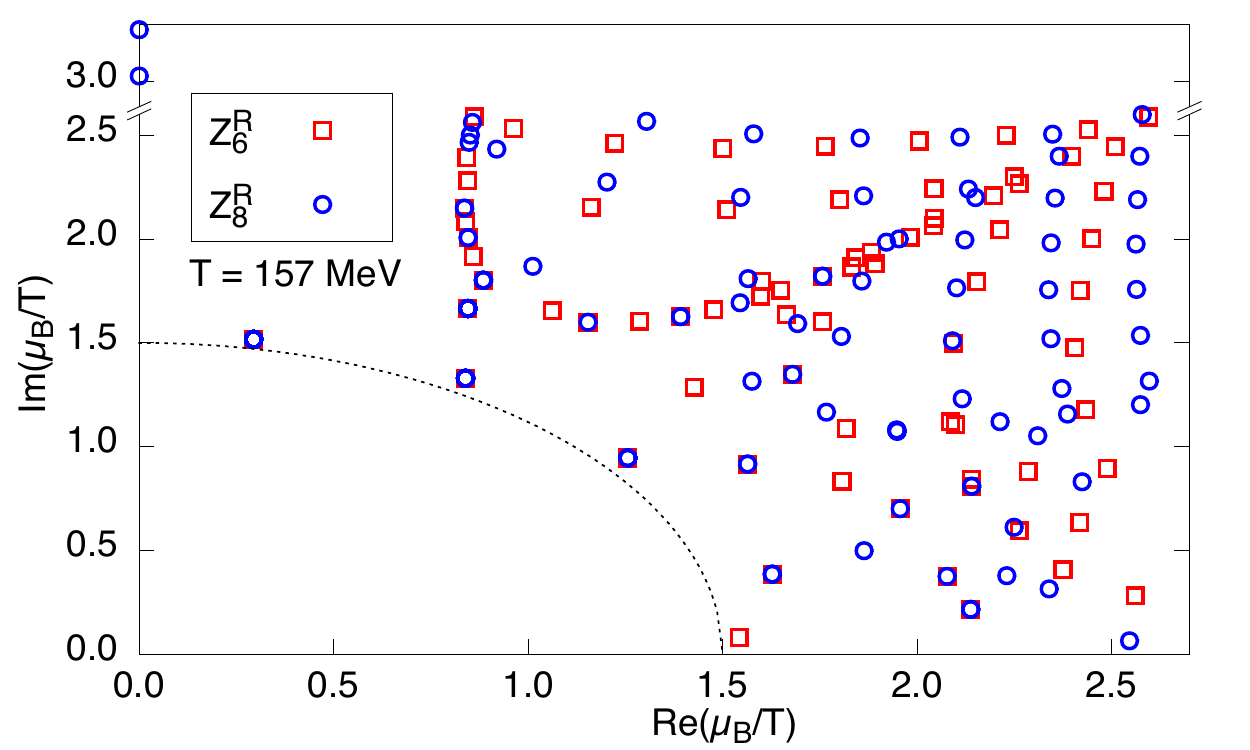}
  \caption{(Top) The average phase factor $\langle\cos\Theta^R_N\rangle$ as a function of $\m/T$. The bands
are the Taylor series expansions of the phase factor to different orders. (Bottom) Zeros of $Z^R_N$
in the complex-$\mu_B$ plane. Only roots in the first quadrant are shown since the distribution
is symmetric in the four quadrants. Both top and bottom plots are for $T= 157$~MeV.}
  \label{fig:phasefac_roots_b6390}
\end{figure}

To investigate the origin of this breakdown, we computed the average phase as a
function of real $\m$, \textit{c.f.} \autoref{eq:theta}. The results are shown in
\autoref{fig:phasefac_roots_b6390} (top). Also, $\av{\cos\thr_N}\approx0$ for
$\m/T\gtrsim1.5$, which shows that the sign problem is uncontrollably severe where
the EoS  calculations broke down. The resummation method thus faithfully captures the
severity of the sign problem, as opposed to the Taylor expansion. The phase
factor cannot be calculated exactly within the Taylor series approach. Its Taylor
series expansion too converges very slowly, as the bands plotted in
\autoref{fig:phasefac_roots_b6390} (top) show. Further, we
searched for the zeros of resummed partition function, \textit{c.f.} \autoref{eq:Z},
in the complex-$\m$ plane. We solved for $\zr_N=0$ using the Newton-Raphson algorithm
with initial guesses chosen from a uniform distribution over a grid
$0\leq\{\Re(\m/T),\Im(\m/T)\}\leq2.5$. The results are shown in
\autoref{fig:phasefac_roots_b6390} (bottom). The zeros of $\zr_6$ and $\zr_8$ are
more or less consistent with each other and appears only for $\abs{\m/T}\gtrsim1.5$.
The exact nature of the singularity responsible for breakdown of the resummation
method is certainly of great interest, \textit{i.e.} whether it is associated with
the Yang-Lee edge singularity of the QCD chiral transition~\cite{Stephanov:2006dn,
Mukherjee:2019eou} or the QCD critical point and approaches the real
axis~\cite{Datta:2016ukp, Gupta:2014qka, Gavai:2008zr, Fodor:2001au, Fodor:2004nz}
\textit{etc.} This will need detailed quantitative studies involving careful
finite-volume scaling analyses using more sophisticated
techniques~\cite{Barbour:1991vs, Wakayama:2018wkc, Giordano:2019slo} and is beyond
the scope of the present work. But our results demonstrate that the breakdown of the
resummation method reflects the associated singularities of the partition function,
at least qualitatively.

\begin{figure}[!t]
  \centering
  \includegraphics[width=0.4\textwidth]{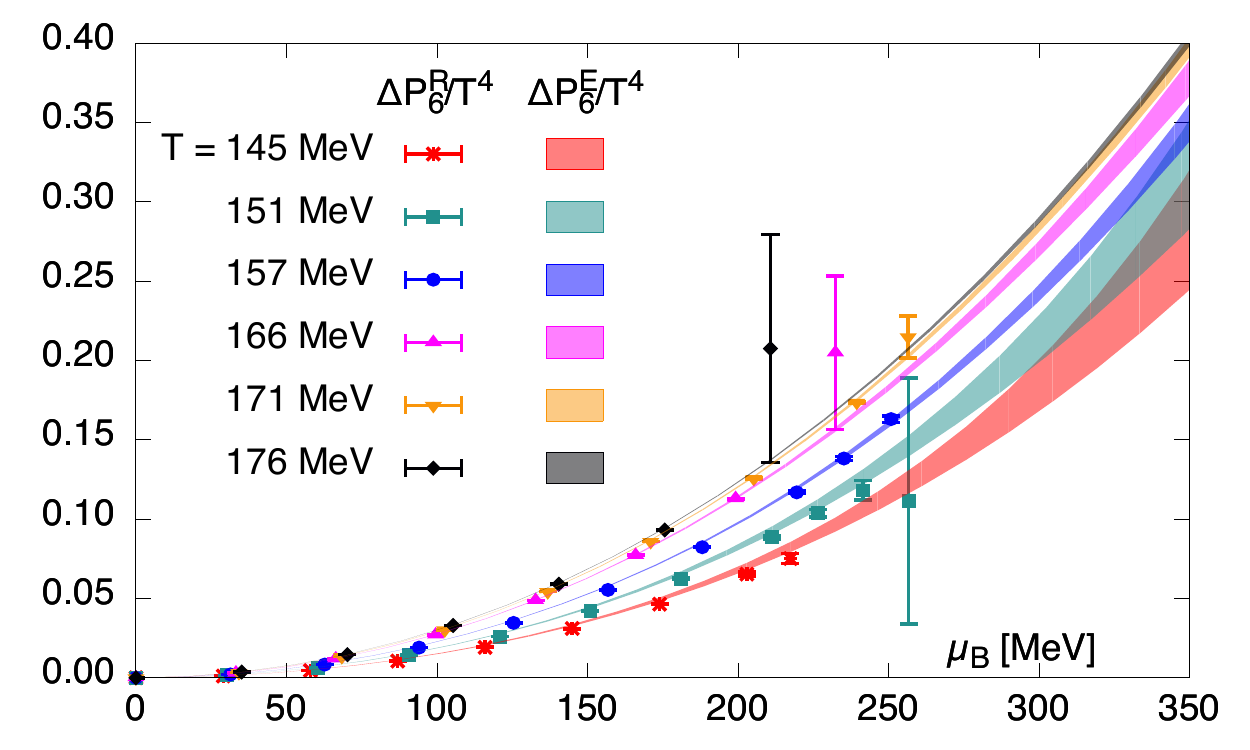}
  \includegraphics[width=0.4\textwidth]{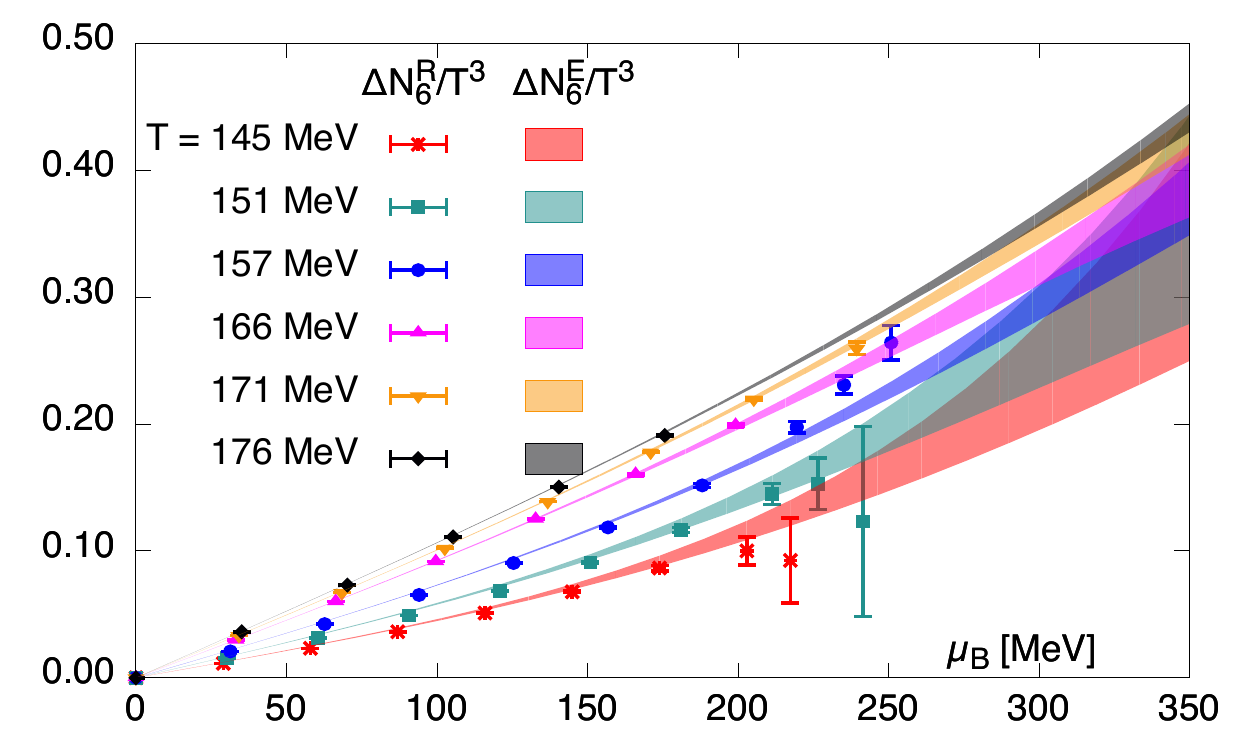}
  \caption{Comparisons between the excess pressure (top) and the net baryon-number
  density (bottom) obtained the sixth order resummation ($\pr_6$ and $\ndr_6$) and
  Taylor expansion ($\pt_6$ and $\ndt_6$) methods for all six temperatures that
  were considered in this work.}
  \label{fig:results_allT}
\end{figure}

Finally, in \autoref{fig:results_allT} we summarize results for all $T=145-176$~MeV
by showing  comparisons between $\pr_6$ and $\ndr_6$ with the corresponding $\pt_6$
and $\ndt_6$.  As in the case of $T=157$~MeV,  $\pr$ and $\ndr$ show improved
convergence over $\pt$ and $\ndt$ at all temperatures. Again, in contrast to the
Taylor expansion the resummation method shows signs of breakdown for
$\m\gtrsim200-250$~MeV,  depending on the temperature. As before, we checked that in
all cases, these breakdowns reflect the severity of the sign problem and the
singularities of the partition function in the complex-$\m$ plane.


\emph{Generalization to multi-parameter and joint expansion in $T, \m$.--} Akin
to multi-parameter reweighting~\cite{Fodor:2001au, Fodor:2004nz, Ejiri:2004yw,
Saito:2013vja} in bare gauge coupling, $\Delta\beta=\beta-\beta_0$, and quark
mass, $\Delta m=m-m_0$, our resummation scheme also can be extended to obtain
$\zr_N(T,\m)$ starting from a different temperature $T_0(\beta_0)$ and bare quark
mass $m_0$,
\begin{align}
  \frac{\zr_N(T,\m)}{Z(T_0,0)} = \left\langle
  e^{-S_G\dbeta + \sum_{i+j=1}^N\Gn{ij} \left(\frac{\m}{T}\right)^i
  \left(\frac{\dm}{T}\right)^j}
  \right\rangle \,,
  \label{eq:rw}
\end{align}
where the expectation value is taken over gauge fields associated with
$\{\beta_0, m_0, 0\}$.  Here, $S_G$ is the pure gauge action and
\begin{align}
  \Gn{ij}(\beta_0,m_0) = \left.
  \frac{\partial^i\partial^j\ln\det[M(m,\m)]}{i!\,j!\,\partial(\m/T)^i\partial(m/T)^j}
  \right\vert_{(m_0,0)} \,.
  \label{eq:G}
\end{align}
Note, $\Gn{i0}=\Dn{i}$ (\autoref{eq:D}), $\Gn{0j}$ are the chiral condensate and
higher order chiral susceptibilities, and general $\Gn{ij}$ are $\m$-derivatives
of various chiral observables~\cite{HotQCD:2018pds, Steinbrecher:2018jbv,
Allton:2005gk, Gupta:2004xs}. This generalization can possibly mitigate the
overlap problem that one might encounter while resumming only in $\m$. Further,
a systematic expansion of the logarithm of \autoref{eq:rw} in powers of
$\dbeta$, $\dm$ and $\m$ yields the expansion of the pressure difference,
$P(T,\m)/T^4-P(T_0,\m)/T_0^4$, in powers of $\dT=T-T_0$ and $\m$; particular
choice of $T_0(\m)$ defined by a line of  constant physics in the $T$-$\m$-plane
reproduces the expansion scheme used in Ref.~\cite{Borsanyi:2021sxv}
by resumming up to $N$-point baryon-current
correlations to all orders in $\m$ and $\dT$~\footnote{See
Appendix for a generalization of the method presented here to resummation in
both $\m$ and $T$. The generalized method resums the alternative expansion scheme presented in S. Borsányi et al. [45]
to all orders in $\m$ and $T$ and is equivalent to multi-parameter
reweighting in $\m$ and $T$}. Thus, our method also generalizes the alternative expansion
scheme of Ref.~\cite{Borsanyi:2021sxv}.

%

\emph{Conclusions.--} We have introduced a new method to compute lattice QCD EoS
by resumming contributions of up to $N$-point baryon-current correlations to all
orders in $\m$. When expanded in powers of $\m$ this resummed partition function
exactly reproduces the Taylor expansion up to $\order{\m^N}$, plus an infinite
series in $\m$ capturing all possible contributions involving only the  $n\le
N$-point baryon-current correlations. This resummation method also amounts to an
approximate reweighting method, thereby bridging two traditional lattice QCD
techniques for $\m\ne0$. With illustrative high-statistics lattice QCD
computations we have demonstrated that the resummation method show improved
convergence over the Taylor expansion method. The method also faithfully
captures the severity of the sign problem as well as reflects the singularities
in the complex-$\m$ plane that are responsible for its eventual breakdown. Thus
the resummation method not only provides a more convergent lattice QCD EoS but
also a more reliable one by enabling us to judge its validity with increasing
$\m$.  Although the resummation method is more general and powerful than the
Taylor expansion, computationally it is somewhat simpler. The resummation method
relies on the computations of $D_n$ which come as an intermediate step in the
computations of the Taylor coefficients. Comparison with the resummed results
and the direct lattice QCD simulations for purely imaginary $\m$ will help us to
decide up to what values $\Im(\m)$ an analytic continuation using only the power
series of $\m$ is justified and whether Pad\'e-type analytic
continuations~\cite{Cea:2012ev, Pasztor:2020dur, Schmidt:2021pey} are necessary
to avoid singularities in the complex-$\m$ plane. We have also introduced a
generalized multi-parameter version of the resummation, \autoref{eq:rw}, and
shown that the method of Ref.~\cite{Borsanyi:2021sxv} is a special case of
this-- Taylor expansion of \autoref{eq:rw} in $T$ and $\m$ along a specific line
in the $T$-$\m$-plane.

%
%
\emph{Acknowledgments.--} We are indebted to members of the HotQCD collaboration for
letting us reuse the data they had generated for the Taylor expansion calculations
as well as for several valuable discussions.

This material is based upon work supported by: (i) The U.S. Department of Energy,
Office of Science, Office of Nuclear Physics through the Contract No.~DE-SC0012704;
(ii) The U.S. Department of Energy, Office of Science, Office of Nuclear Physics and
Office of Advanced Scientific Computing Research, within the framework of Scientific
Discovery through Advance Computing (SciDAC) award Computing the Properties of Matter
with Leadership Computing Resources; (iii) The U.S. Department of Energy, Office of
Science, Office of Nuclear Physics, within the framework of the Beam Energy Scan
Theory (BEST) Topical Collaboration. (iv) The Early Career Research Award
of the Science and Engineering Research Board (SERB) of the
Government of India; and (v) The Institute Postdoc Fellowships of
the Indian Institute of Science, Bangalore.

(i) This research used awards of computer time provided by the INCITE and ALCC
programs at Oak Ridge Leadership Computing Facility, a DOE Office of Science User
Facility operated under Contract No. DE-AC05-00OR22725. (ii) This work also made use
of the clusters and data storage facilities located at Bielefeld University, Germany.

\bibliographystyle{apsrev4-2.bst}
\bibliography{refs}
\begin{appendix}
\section{Generalization to multi-parameter and joint expansion in $\mathbf{\emph{T}}$ and $\pmb{\m}$}

Our starting point for the multi-parameter expansion is \autoref{eq:rw}.
There, the expectation value is taken over gauge fields associated with
$\{\beta_0, m_0, \m=0\}$, where $\beta$ is the QCD gauge coupling, $S_G$ is the pure gauge action and
the $\Gn{ij}$ are given by \autoref{eq:G}.

For brevity, in this section, we will use notations $\hm\equiv\m/T$, $\hms=m/T$
and $\hdm\equiv\hms-\hms_0$, and  provide explicit demonstration of the
generalized resummation by keeping only leading order terms, \textit{i.e.}
$\order{\dbeta}$, $\order{\hdm}$ and $\order{\hm^2}$, in all expansions.
Extensions to higher orders are straightforward. We consider the case where the
temporal extent ($N_\tau$) of the lattice is kept fixed. In this case, $T$ is
changed by varying the bare gauge coupling $\beta$, and the bare quark mass $m$ must
be tuned with $\beta$ to keep vacuum hadron masses constant. Thus, $T(\beta, m)$
and $T_0(\beta_0, m_0)$. Applying chain rule for derivatives as well as
expanding $T(\beta, m)$ around $(\beta_0, m_0)$ one gets
\begin{equation}
\left. \dT\pdv{}{T} \right\vert_{T_0} =
\left. \dbeta\pdv{}{\beta} \right\vert_{\beta_0} +
\left. \hdm\pdv{}{\hms} \right\vert_{\hms_0} + \dots \,.
\label{eq:pdvT}
\end{equation}
With
\begin{align}
  \langle \Dn{i}^j \rangle = \frac{1}{Z(\beta,m)}
  \int \mathcal{D}U\; \Dn{i}^j\; e^{-\beta S_G[U]+\ln\det M(m)} \,,
\end{align}
to the leading order
\begin{align}
\dT\,\frac{d\langle\Dn{i}^j\rangle}{dT} =
& -\left[\langle S_G \Dn{i}^j \rangle -
\langle S_G\rangle \langle\Dn{i}^j\rangle\right]\dbeta \notag \\
& + \left[\langle\Gn{01}\Dn{i}^j \rangle
 - \langle \Gn{01} \rangle\langle \Dn{i}^j \rangle
 + j\langle \Dn{i}^{j-1}\Gn{i1}\rangle \right] \hdm  \notag \\
& + \dots \,.
\label{eq:dDdT}
\end{align}

The goal is to obtain $\zr_N(T,\m)$ by expanding around $Z(T_0,0)$, while
resumming contributions of up to $N$-point baryon-current correlations to all
orders in $\m$ and $\dT$. Following multi-parameter reweighting technique
\begin{equation}
\frac{Z(T,\m)}{Z(T_0,0)} = \left\langle
e^{-\dbeta S_G}\frac{\det M(m,\m)}{\det M(m_0,0)}\right\rangle \,,
\label{eq:rwZ}
\end{equation}
where the expectation value is with respect to a gauge field ensemble generated
for $\{\beta_0,m_0,\m=0\}$. Simultaneously expanding in $\m$ and $\dm$
the determinant
ratio in \autoref{eq:rwZ} can be written as
\begin{equation}
\frac{\det M(m,\m)}{\det M(m_0,0)} = \exp\left[
\sum_{i+j=1}^\infty\Gn{ij} \hm^i \hms^j
\right] \,,
\label{eq:detratio}
\end{equation}
where $\Gn{ij}$ are defined through \autoref{eq:G}. By plugging \autoref{eq:detratio} back into \autoref{eq:rwZ} 
and truncating the sum at $i+j=N$, we obtain \autoref{eq:rw}.

Next, by Taylor expanding \autoref{eq:rw} in powers of $\dbeta$, $\hdm$ and $\hm$ we get
\begin{align}
\frac{Z(T,\m)}{Z(T_0,0)} & = 1 -
\langle S_G \rangle \dbeta + \langle \Gn{01} \rangle \hdm + \langle\Dn{1}\rangle\hm \notag \\
& - \left[\langle S_G\Dn{1}\rangle\dbeta - \langle\Gn{01}\Dn{1}+\Gn{11}\rangle\hdm\right]\hm \notag \\
& + \langle \Dn{2} + \Dn{1}^2/2 \rangle \hm^2 \notag \\
& - \left[ \langle S_G (\Dn{2}+\Dn{1}^2/2)\rangle \dbeta - \langle \Gn{01}(\Dn{2}+\Dn{1}^2)/2\rangle\hdm \right. \notag \\
& \left. -\langle\Gn{21}+\Gn{11}\Dn{1}\rangle\dm \right] \hm^2 + \;\dots \,.
\label{eq:Zexp}
\end{align}
The pressure difference is given by
\begin{align}
  \Delta\left[\frac{P}{T^4}\right] \equiv
  \frac{P(T,\m)}{T^4} - \frac{P(T_0,0)}{T_0^4} =
  \frac{N_\tau^3}{N_s^3} \ln\left[\frac{Z(T,\m)}{Z(T_0,0)}\right] \,,
  \notag
\end{align}
where $N_s$ is the spatial extent of the lattice. Using \autoref{eq:Zexp},
expanding the logarithm in powers of $\dbeta$, $\hdm$, $\hm$ and keeping only
the real part one obtains
\begin{align}
  \frac{N_s^3}{N_\tau^3} & \Delta\left[\frac{P}{T^4}\right] =
  \langle \Gn{01} \rangle \hdm -\langle S_G \rangle \dbeta
  + \langle\Dn{2}+\Dn{1}^2/2\rangle \hm^2 \notag \\
  & - \left[ \langle S_G(\Dn{2}+\Dn{1}^2/2)\rangle
  - \langle S_G \rangle \langle (\Dn{2}+\Dn{1}^2/2) \rangle \right] \hm^2\dbeta \notag \\
  & + \left[ \langle \Gn{01}(\Dn{2}+\Dn{1}/2)\rangle
  - \langle \Gn{01} \rangle\langle(\Dn{2}+\Dn{1}^2/2)\rangle \right. \notag \\
  & \left. + \langle\Gn{21}+\Gn{11}\Dn{1} \rangle \right] \hm^2\hdm + \dots \,.
  \label{eq:Pexp1}
\end{align}
Noting that
\begin{align}
  \left. \frac{d[P(T,0)/T^4]}{dT} \right\vert_{T_0} \dT =
  \langle \Gn{01} \rangle \hdm -\langle S_G \rangle \dbeta \,,
\end{align}
and using \autoref{eq:pdvT} of the main paper, it is easy to
identify that \autoref{eq:Pexp1} is nothing but a joint Taylor expansion of
$P(T,\m)$ in $T$ and $\m$ around $(T_0,0)$,
\begin{align}
  \Delta\left[\frac{P}{T^4}\right] & =
  \left. \frac{d[P(T,0)/T^4]}{dT} \right\vert_{T_0} \dT
  + \frac{1}{2!}\cb_2(T_0)\hm^2 \notag \\
  & + \frac{1}{2!} \left. \frac{d\cb_2(T)}{dT} \right\vert_{T_0} \hm^2\dT
  + \order{\hm^4, (\dT)^2}\,.
  \label{eq:Pexp2}
\end{align}
Thus, the generalized version given by \autoref{eq:rw} genuinely resums
contributions of up to $N$-point baryon current in the Taylor expansion of EoS
to all orders in $T,\m$.

Following Ref.~\cite{Borsanyi:2021sxv}, 
the generalized resummation of
\autoref{eq:rw} can be made even more powerful by choosing the expansion point
$T_0$ along some physically motivated line in the $T$-$\m$-plane, \textit{i.e.}
by choosing some physically motivated $\beta_0(\m)$ and $m_0(\m)$. The
one-to-one correspondence between the Taylor expansion of \autoref{eq:rw} and
alternative expansion scheme presented in Ref.~\cite{Borsanyi:2021sxv} can be readily observed. By
including the $\mathcal{O}(\hm^4)$ in \autoref{eq:Pexp2} and taking a
$\hm$-derivative we get
\begin{align}
  \cb_1(T,\m) & = \hm\cb_2(T_0,0)
  + \left. \frac{d\cb_2(T,0)}{dT} \right\vert_{T_0} \hm\dT  \notag \\
  & + \frac{1}{6}\cb_4(T_0,0)\hm^3 + \dots \,.
  \label{eq:chi1}
\end{align}
If one chooses
\begin{align}
  T_0(\hm) = T - \frac{1}{6}\frac{\cb_4(T_0,0)}{(d\cb_2(T,0)/dT)_{T_0}}\hm^2 \,,
\end{align}
in \autoref{eq:chi1} then one arrives at the starting point of
Ref.~\cite{Borsanyi:2021sxv}, namely $\cb_1(T_0,\hm) = \hm \cb_2(T_0,0)$. Hence,
the method used in  Ref.~\cite{Borsanyi:2021sxv} is a special Taylor-expanded
case of the generalized resummation \autoref{eq:rw}.
\end{appendix}
\end{document}